# Localized excitation in the hybridization gap in YbAl$_3$

A. D. Christianson,[1,2] J. M. Lawrence,[1] E. A. Goremychkin,[3,4] R. Osborn,[3] E. D. Bauer,[2] J. L. Sarrao,[2] J. D. Thompson,[2] C. D. Frost,[4] and J. L. Zarestky[5]

[1]University of California, Irvine, California 92697, USA

[2] Los Alamos National Laboratory, Los Alamos, New Mexico 87545, USA

[3] Argonne National Laboratory, Argonne, Illinois 60439, USA

[4]ISIS Facility, Rutherford Appleton Laboratory, Chilton, Didcot OX11 0QX, UK

[5]Ames Laboratory, Iowa State University, Ames, Iowa 50011, USA

The intermediate valence compound YbAl$_3$ exhibits a broad magnetic excitation in the inelastic neutron scattering spectrum with characteristic energy $E_1 \approx 50$ meV, equal to the Kondo energy ($T_K \sim 600\text{-}700$ K). In the low temperature ($T < T_{coh} \sim 40$ K) Fermi liquid state, however, a new peak in the scattering occurs at $E_2 \approx 33$ meV, which lies in the hybridization gap that exists in this compound. We show, using inelastic neutron scattering on a single-crystal sample, that while the scattering at energies near $E_1$ has the momentum ($Q$-) dependence expected for interband scattering across the indirect gap, the scattering near $E_2$ is independent of $Q$ over a large fraction of the Brillouin zone. A possible explanation is that the peak at $E_2$ arises from a spatially-localized excitation in the hybridization gap.

61.12.Ex, 71.28.+d, 75.20.Hr

The Anderson Lattice,[1,2] in which a small (~ 10-30 meV) semiconducting gap arises from hybridization of the 4f electrons with the conduction electrons (Fig. 1f), is believed to capture the essence of the physics of intermediate valence compounds. When the Fermi level is in the gap, the behavior is that of the Kondo insulators such as $YbB_{12}$ and $SmB_6$. The most direct experimental evidence for the gap comes from measurements of the optical conductivity. For example, for $YbB_{12}$ the low temperature conductivity vanishes below 25 meV, and then rises to a maximum in the mid-IR near 0.25 eV.[3] The experimental neutron scattering spectra of the Kondo insulators also show low temperature features whose energy scale is that of the gap. The intensity of these features is highly $Q$-dependent, peaking at the (1/2,1/2,1/2) zone boundary point in $YbB_{12}$[4] and $SmB_6$.[5]

When the Fermi level is not in the gap, the behavior is that of an intermediate valence metal. The cubic ($Cu_3Au$ structure) compound $YbAl_3$ exhibits strong intermediate valence ($z$ = 2.75) and a large Kondo temperature ($T_K$ ~ 600-700 K). Below the "coherence temperature" $T_{coh}$ ~ 40 K, where the resistivity exhibits the $T^2$ behavior expected for a Fermi liquid,[6] the optical conductivity exhibits[7] a narrow Drude resonance separated by a deep minimum at 30 meV from a mid-infrared peak at 0.25 eV. It is believed that the minimum and the mid-IR peak arise from the (vertical) interband transitions across the hybridization gap, while the Drude resonance arises from intraband transitions across the Fermi level, which lies in a region of high density of states near the zone boundary. Inelastic neutron scattering (INS) experiments[8] in polycrystalline samples of $YbAl_3$ have shown that for $T$ > 40 K the magnetic scattering is broad with a characteristic energy of order $E_1 \approx$ 50 meV, corresponding to the Kondo temperature, but at low temperature an additional narrow peak occurs in the INS near $E_2 = 33$ meV. This is the same energy as the minimum in the optical conductivity, i.e it occurs on the same scale as the hybridization gap. This peak broadens and weakens on alloying with a small concentration of Lu[9] which fact, taken together with the disappearance of the peak above 40 K, means it is a property of the fully coherent ground state. The characterization of this scattering and the relationship of the behavior to that seen in the Kondo insulators is an important goal.

    To explore the physics of this scattering, and in particular to determine the *Q*-dependence, we have performed inelastic neutron scattering measurements on single crystals of YbAl$_3$. The crystals were grown by precipitation from excess aluminum (self-flux method). One set of experiments were performed at 6 K and 100 K on the MAPS time-of-flight spectrometer at the ISIS Pulsed Neutron and Muon Facility of the Rutherford Appleton Laboratory. The initial energy was $E_i = 120$ meV. Four crystals, of total mass ~5 g, were mounted on an aluminum sample holder and co-aligned with a mosaic of 2.5º. We set *k$_i$* (the incoming beam wavevector) initially parallel to the [1,0,0] direction; in a second set of measurements we chose *k$_i$* // [1,1,0]. MAPS employs a large pixellated detector where each individual pixel element detects neutrons on a trajectory through reciprocal space parameterised by the incident energy, $E_i$, the angles θ and φ between the incident *k$_i$*, and final wavevector *k$_f$* and the time-of-flight or energy transfer Δ*E*. The amalgamation of these simultaneous individual trajectories results in a 3D hypersurface in the 4D reciprocal space of *Q* and Δ*E*. Unlike the case of a triple-axis spectrometer where the crystal is rotated to construct a scan over energy transfer with *Q* =

$(2\pi/a_0)$ $(h, k\ l)$ held constant, for MAPS, where the crystal orientation is fixed during the scan, only three of the four variables $h$, $k$, $l$ and $\Delta E$ are independent in any plot at fixed $E_i$.

In Fig. 1 we plot the scattering at $T = 6$ K as a function of energy transfer at different momentum transfers. In agreement with the earlier results reported for polycrystals,[8,9] the scattering shows two features: a peak centered at $E_2 = 33$ meV, whose width is essentially equal to the instrumental resolution, and a broad feature centered near $E_1 = 50$ meV. In Figs. 1a-d, both $h$ and $k$ vary, with $h + k$ constant, as $\Delta E$ varies; in Fig. 1e, only $h$ varies with $\Delta E$. The values of $(h,k,l)$ given in the figure correspond to the values at $E_1$ and at $E_2$. The scattering near $E_1$ shows considerable variation of intensity and lineshape with $Q$, but the peak near $E_2$ does not vary much with $Q$. Since YbAl$_3$ has the Cu$_3$Au crystal structure (with $a_0 = 4.203$ Å), the Brillouin zone (BZ) is a simple cube centered at $\Gamma$ and extending ±0.5 in reduced $(h,k\ l)$ units in all three directions. Hence, Fig. 1 shows the $E_2$ peak at key positions in the BZ, including the $\Gamma$ point (Fig. 1a), various zone boundary points (Figs. 1b, d and e) and a point in the middle of the BZ (Fig. 1c). Plots at other positions in the BZ are very similar. The variation in the position and the magnitude of this peak is small, suggesting that the peak at $E_2$ is essentially independent of $Q$.

Fig. 2a shows an intensity map for the projection onto the $k$-$l$ plane for an interval of energy transfer $E_1 \pm 5$ meV. In this plot, the reduced wavevector $h$ varies with $k$ and $l$, e.g. for $l = 0$ it varies from 1.2 at $k = 0$ to 1.4 at $k = 1.5$. The cubic symmetry is readily apparent in this plot. Peaks are observed near $(h,0.5,0.5)$ where $h(\Delta E = E_1) = 1.2$. These peaks lie essentially at the zone boundary close to the [0,1,1] direction. Fig. 2b shows an intensity plot in the $\Delta E$-$Q_k$ plane for $-0.2 < l < 0.2$. For $28 < \Delta E < 38$ meV, $h$ varies from 0.85 at $k = 0$ to 1.05 at $k = 1.5$, crossing the zone boundary at $k = 0.5$ and 1.5. This scattering for $\Delta E = E_2 \pm 5$ meV is basically independent of $Q_k$.

Preliminary work (not shown here) on the PHAROS time-of-flight spectrometer at the Los Alamos Neutron Science Center (LANSCE) showed that the magnetic scattering in Figs.1 and 2 overlaps with phonon scattering. To determine each contribution requires larger momentum transfer than is available on MAPS. To this end, we have measured the spectra on the HB3 triple-axis spectrometer at the High Flux Isotope Reactor (HFIR) at Oak Ridge National Laboratory. Six flux-grown crystals, each approximately 1 g, were coaligned to within ±0.3° with the [1,-1,0] direction vertical. We measured the spectra at 10 and 100 K, using a fixed final energy of 14.7 meV and a collimation 30'-40'-80'-240'. The counting rate is normalized to monitor count units (1 mcu ≈ 1 sec). For each value of $\mathbf{Q}_1 = (2\pi/a_0)(h,k\,l)$ we measured the 10 K spectra at a second momentum transfer $\mathbf{Q}_2 = (2\pi/a_0)(h,k\,l+2)$ that is equivalent to $\mathbf{Q}_1$ in the reduced zone scheme and has a similar structure factor, and then fit the data as the sum of a magnetic contribution $M(E)$ that varies with $Q$ as the 4$f$ form factor, a phonon contribution $Ph(E)$ that scales as $Q^2$ and a constant background, measured in energy gain. An example is shown in Fig. 3a and b, where it can be seen that the phonon contribution is small at low $Q$, as expected. As a check on our procedure, we subtracted $Ph(E)$ (scaled by the Boson temperature factor) from the data at 100 K, finding that the resulting magnetic contribution $M$ at 100 K has the form shown in Fig. 3c, independent of (reduced) $Q$. This spectrum is very similar to that seen in polycrystals[8] at 100 K; the sharp feature at $E_2$ disappears above 40 K. The $Q$-independence suggests that the spin dynamics for $T > T_{coh}$ is that of incoherent Kondo scattering.

Fig. 4 shows the magnetic contribution $M$ at 10 K with the 4$f$ form factor divided out in order to represent the scattering in the reduced zone. The scattering is essentially identical at zone center, and at the (0,0,1/2) and (1/2,1/2,0) zone boundary points. This confirms our time-of-flight result that the 30 meV scattering is largely $Q$-independent. We note, however, that the scattering at the (1/2,1/2,1/2) zone boundary point is considerably weaker. This effect can also be seen in the time-of-flight data in plots (not shown here) similar to Fig. 2, so we believe it is a real effect.

As seen in Figs. 1, 2 and 4, our basic result is that the scattering near $E_1 = 50$ meV is highly $Q$-dependent, with peaks at the (1/2,1/2,0) zone boundary, but apart from the variation with the 4$f$ form factor, the scattering near $E_2 = 33$ meV appears to be independent of $Q$ over an appreciable fraction of the Brillouin zone.

The neutron scattering intensity is proportional to

$$\int N_i(\Sigma_q) f(\Sigma_q) N_f(\Sigma_{q-Q}) f(\Sigma_{q-Q}) d\Sigma_q$$

where $\Sigma_{q-Q} - \Sigma_q = \Delta E$ is the energy transfer, $Q$ is the momentum transfer, $f$ is the Fermi function and $N_i$ and $N_f$ are the initial and final densities of states. In calculations[1,2] of the interband transitions in the renormalized band structure (Fig. 1f) of the Anderson lattice, such scattering is highly $Q$-dependent. The most intense scattering occurs when the energy transfer $\Delta E$ equals the threshold for indirect transitions between the regions of large density of states at the zone center and zone boundary of the upper and lower bands respectively. This occurs when $Q = q_{BZ}$ (Fig. 1f). At smaller $Q$, the scattering is weaker and occurs at higher energy. When the Fermi level does not lie in the gap, so that the behavior is that of an intermediate valence (IV) metal, the interband transitions still occur, and show a very similar Q-dependence to that of the Kondo insulators (See Fig. 5 of Ref. 1). In addition, low energy intraband scattering across the Fermi surface is expected; this is related to the Drude scattering in the optical conductivity. The characteristic energy for such scattering varies linearly with $Q$, as expected on general grounds for a Fermi liquid.

Sharp excitations at low temperatures, whose energy scale is that of the hybridization gap, have been observed at low temperatures in the Kondo insulators $SmB_6$,[5] $YbB_{12}$[4] and TmSe.[10] These excitations are highly $Q$-dependent. For example, in TmSe the intensity is at least a factor of four smaller at zone center than at the zone boundary, and in $YbB_{12}$ and $SmB_6$ the intensity is largest at the zone boundary along the [1,1,1] direction, decreasing dramatically as $Q$ increases or as the angle of $Q$ with respect to the [1,1,1] direction changes. The scattering thus has the behavior expected for threshold scattering at the zone boundary.

In $YbAl_3$, the fact that the scattering near $E_1 = 50$meV shows considerable $Q$-dependence, peaking at the (1/2,1/2,0) zone boundary point, is consistent with interband scattering across an indirect gap of order 50 meV. We note that this is also the Kondo scale for this compound; if so, the high temperature Kondo scattering evolves into interband scattering with the same energy scale.

However, the $E_2$ peak shows little variation in intensity for different values of $Q$ that include the zone center, key zone boundary points and points in the center of the reduced zone (Figs. 1 and 4); and in addition the scattering is *weaker* at the (1/2,1/2,1/2) point (Fig. 4). Hence this excitation differs in very important respects from those seen in the Kondo insulators. It clearly does not have the $Q$-dependence expected for threshold interband scattering, unless the actual renormalized band structure is very different from that shown in Fig. 1f. This excitation occurs at too large an energy, and lacks the expected $Q$-dependence, to arise from intraband Fermi surface scattering. Since the amplitude of this peak is strongly diminished at 100K, it cannot represent a crystal field transition; and, indeed, well-defined crystal field excitations are not expected in intermediate valence systems with such large Kondo temperatures.

An important possibility is that the scattering represents a spatially localized excitation. As mentioned above, the energy of the excitation coincides with the deep minimum in the optical conductivity; this suggests that the localized excitation lies inside the hybridization gap, thus having the character of an exciton. Riseborough[11] showed how a magnetic bound state, whose energy lies below the interband threshold, can arise in the Kondo insulators due to RKKY interactions that are enhanced by the proximity of the $f$-level to the Fermi surface. Such a bound state, however, should be large for zone-boundary $Q$, dispersing and losing amplitude rapidly as $Q$ decreases. Brandow[2] postulated the existence of a localized excitation in the form of an onsite valence fluctuation from the hybridized state into the unhybridized trivalent ($4f^{13}$) state. He showed that if such an elementary excitation, which unbinds or unscreens a local moment at an arbitrary lattice site, is included in a consistent manner with the usual excited states of the Anderson lattice, a two-peak structure of the susceptibility and specific heat should occur, similar to the low temperature anomalies observed[6] experimentally in YbAl$_3$. In his treatment, however, the excitation was not derived from the Anderson lattice Hamiltonian but was included in an *ad hoc* manner. It is thus open question whether and how such an excitation can arise in the context of the Anderson lattice. It is also unclear why a local excitation should be so sensitive to alloy disorder.[9]

We are grateful to Steve Shapiro for sharing and discussing his unpublished data on this compound, and to Peter Riseborough and Steve Nagler for helpful discussions. Work at UC Irvine was supported by the US Department of Energy (DOE) under Grant No. DE-FG03-03ER46036. Work at Argonne National Laboratory was supported by the DOE under Contract No. W-31-109-ENG-38. Ames Laboratory is operated by Iowa State University for the U.S. Department of Energy (DOE) under Contract No. W-7405-ENG-82. Work performed at the HFIR Center for Neutron Scattering was supported by the DOE Office of Basic Energy Sciences Materials Science, under Contract No. DE-AC05-000R22725 with UT-Battelle, LLC. Work at Los Alamos National Laboratory, including at LANSCE, was also performed under the auspices of the DOE.

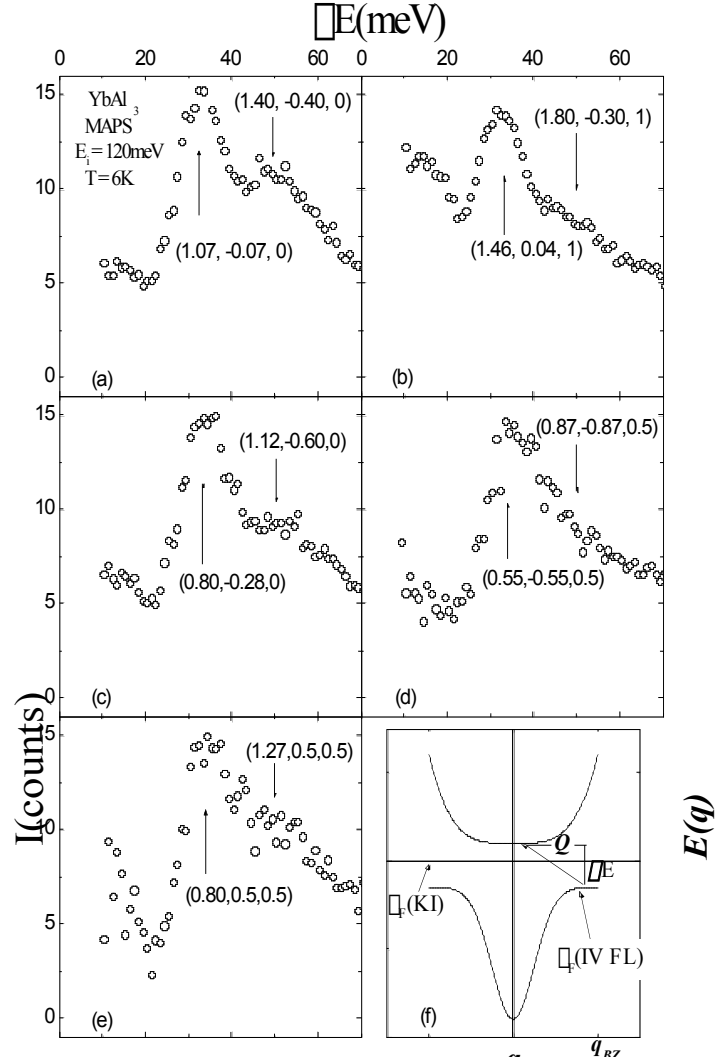

Fig. 1 a-e) The scattering intensity at $T = 6$ K in YbAl$_3$ versus energy transfer at various momentum transfers, measured on MAPS using an initial energy of 120meV. The reduced wavevectors given in each panel correspond to the values at $E_1 = 50$ meV and $E_2 = 33$ meV. Fig. 1f shows a schematic plot of the renormalized band structure expected for the Anderson lattice, with a hybridization gap. For IV Fermi liquids (FL), the Fermi level lies in the high density of states region of the lower band; for Kondo insulators (KI), it falls in the gap. A typical interband transition, with energy transfer $\Delta E$ and momentum transfer $Q$, is shown.

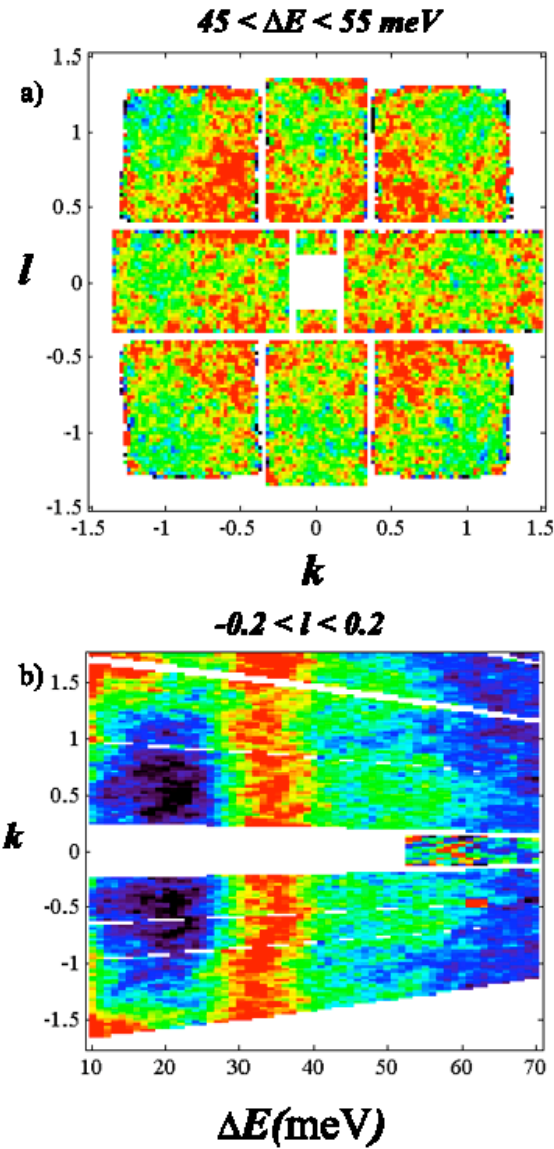

Fig. 2  a) Intensity versus reduced wavevector (*k, l*) for energy transfers in the range 45 < $\Delta E$ < 55 meV for neutron scattering data taken on YbAl$_3$ at *T* = 6 K, using MAPS with $E_i$ = 120 meV.  b) Intensity versus energy transfer $\Delta E$ and reduced wavevector *k* for the component *l* of reduced wavevector in the range -0.2 < *k* < 0.2.

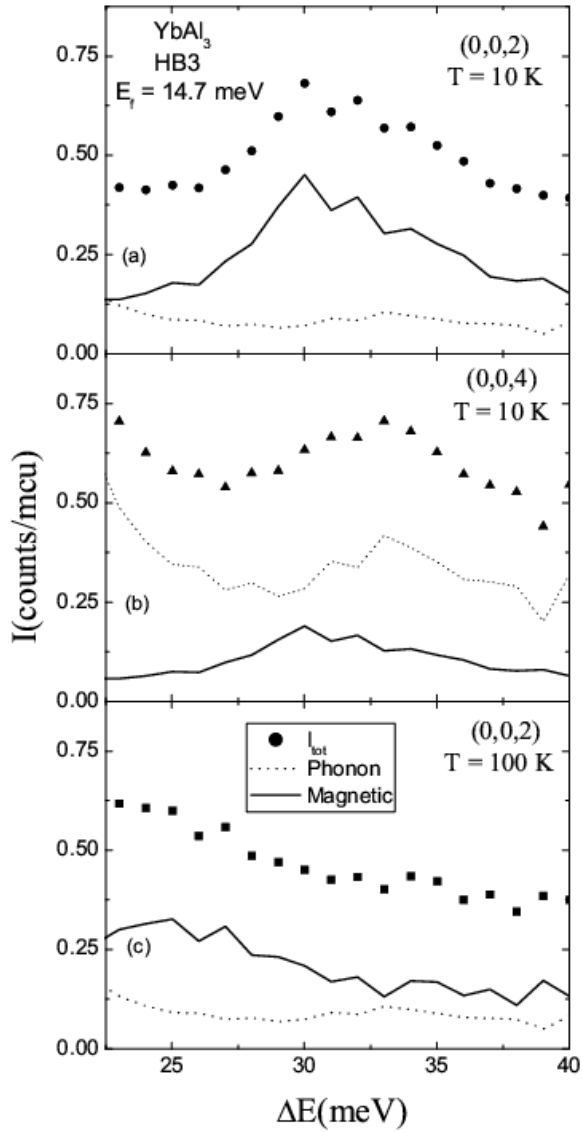

Fig. 3 Intensity, measured relative the monitor counts (1 mcu ≈ 1 sec), versus energy transfer for neutron scattering data taken on YbAl$_3$ using HB3 at $E_f$ = 14.7 meV. Data: solid symbols; magnetic contribution: solid lines; phonon: dotted lines. The background is 0.16 counts/mcu.

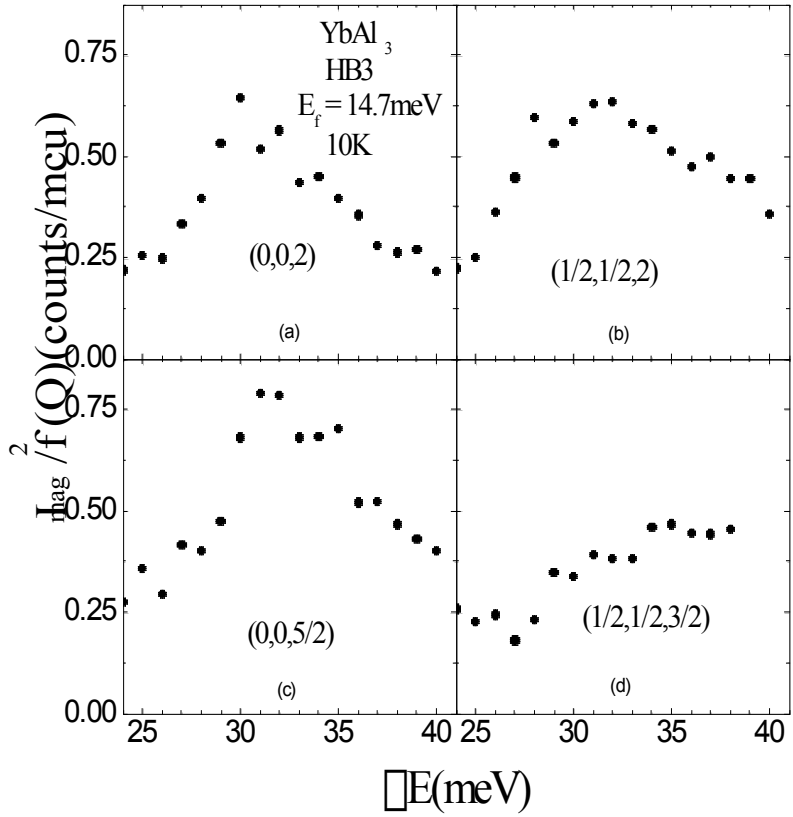

Fig. 4  The magnetic contribution (see text) to the scattering of YbAl$_3$ at $T = 10$ K and at four values of momentum transfer $Q = 2\pi/a_0\ (h,k,l)$. The data had been divided by the 4$f$ form factor, and thus represent the scattering in the reduced zone.

[1] A. A. Aligia and B. Alascio, J. Magn. Magn. Mater. **46**, 321 (1985).

[2] B. H. Brandow, Phys. Rev. B **37**, 250 (1988).

[3] H. Okamura, S. Kimura, H. Shinozaki, T. Nanba, F. Iga, N. Shimizu, and T. Takabatake, Phys. Rev. B **58**, R7496 (1998).

[4] J.-M. Mignot, P. A. Alekseev, K. S. Nemkovski, L.-P. Regnault, F. Iga, and T. Takabatake, Phys. Rev. Lett. **94**, 247204 (2005).

[5] P. A. Alekseev, J.-M. Mignot, J. Rossat-Mignod, V. N. Lazukov, I. P. Sadikov, E. S. Konavalova, and Yu. B. Paderno, J. Phys.: Condens. Matter **7**, 289 (1995).

[6] A. L. Cornelius, J. M. Lawrence, T. Ebihara, P. S. Riseborough, C. H. Booth, M. F. Hundley, P. G. Pagliuso, J. L. Sarrao, J. D. Thompson, M. H. Jung, A. H. Lacerda, and G. H. Kwei, Phys. Rev. Lett. **88**, 117201 (2002).

[7] H. Okamura, T. Michizawa, T. Nanba, and T. Ebihara, J. Phys. Soc. Jpn. **73**, 2045 (2004).

[8] A. P. Murani, Phys. Rev. B **50**, 9882 (1994).

[9] R. Osborn, E. A. Goremychkin, I. L. Sashin, and A. P. Murani, J. Appl. Phys. **85**, 5344 (1999).

[10] S. M. Shapiro and B. H. Grier, Phys. Rev. B **25**, 1457 (1982).

[11] P. S. Riseborough, Ann. Phys. (Leipzig) **9**, 813 (2000).